\documentclass{revtex4}
\usepackage{epsfig} 
\usepackage{epstopdf}
\usepackage{amssymb}
\usepackage{amsmath}
\usepackage{amsfonts}
\usepackage{graphicx}
\graphicspath{ {plots/} }
\usepackage{mathrsfs}
\usepackage{color}
\usepackage{multirow}
\usepackage{psfrag}

\newcommand{\R}{\mathbb{R}}

\newcommand{\E}{\mathbb{E}}

\newcommand{\fc}{\mathfrak{c}}

\newcommand{\ff}{\mathfrak{f}}

\newcommand{\fp}{\mathfrak{p}}

\newcommand{\fz}{\mathfrak{z}}

\newcommand{\bg}{\mathbf{g}}

\newcommand{\cM}{\mathcal{M}}

\newcommand{\cO}{\mathcal{O}}

\newcommand{\cS}{\mathcal{S}}

\newcommand{\cX}{\mathcal{X}}
\newcommand{\cY}{\mathcal{Y}}
\newcommand{\cZ}{\mathcal{Z}}
\newcommand{\be}{\begin{equation}}
\newcommand{\ee}{\end{equation}}
\newcommand{\bea}{\begin{eqnarray}}
\newcommand{\eea}{\end{eqnarray}}
\newcommand{\nn}{\nonumber}
\newcommand{\kt}{\rangle}

\newcommand{\ed}{\end{document}}

\newcommand{\bi}{\begin{itemize}}
\newcommand{\ei}{\end{itemize}}

\newcommand{\bce}{\begin{center}}
\newcommand{\ece}{\end{center}}

\newcommand{\sD}{\mathscr{D}}

\newcommand{\df}{\dot{f}}
\newcommand{\ddf}{\ddot{f}}
\newcommand{\dG}{\dot{G}}
\newcommand{\da}{\dot{a}}
\newcommand{\dda}{\ddot{a}}
\newcommand{\db}{\dot{b}}
\newcommand{\ddb}{\ddot{b}}

\newcommand{\beq}{\begin{equation}}
\newcommand{\eq}{\end{equation}}
\newcommand{\ea}{\end{align}}

\begin{document}

\title{Scattering due to geometry: The case of a spinless particle moving\\ on an asymptotically flat embedded surface}

\author{Neslihan Oflaz$^1$, Ali~Mostafazadeh$^{1}$, and Mehrdad Ahmady$^2$\\[6pt]
$^1$~Departments of Mathematics and Physics, Ko\c{c} University, 34450 Sar{\i}yer,
Istanbul, Turkey\\[3pt]
$^2$~Department of Physics, Azarbaijan University of Shahid Madani,
53714-161 Tabriz, Iran~~
}

\begin{abstract}
A nonrelativistic quantum mechanical particle moving freely on a curved surface feels the effect of the nontrivial geometry of the surface through the kinetic part of the Hamiltonian, which is proportional to the Laplace-Beltrami operator, and a geometric potential, which is a linear combination of the mean and Gaussian curvatures of the surface. The coefficients of these terms cannot be uniquely determined by general principles of quantum mechanics but enter the calculation of various physical quantities. We examine their contribution to the geometric scattering of a scalar particle moving on
an asymptotically flat embedded surface. In particular, having in mind the possibility of an experimental realization of the geometric scattering in a low density electron gas formed on a bumped surface, we determine the scattering amplitude for arbitrary choices of the curvature coefficients for a surface with global or local cylindrical symmetry. We also examine the effect of perturbations that violate this symmetry and consider surfaces involving bumps that form a lattice.


\end{abstract}

\maketitle

\section{Introduction}

The study of quantum mechanics of nonrelativisitic  particles moving
in a curved Riemannian manifold has been a focus of attention since
the early days of canonical quantum gravity. In his pioneering works
of the 1950's, Bryce DeWitt explored the consequences of the
form-invariance of the Hamiltonian under the group of general point
transformations of the configuration space \cite{DeWitt-1952} and
discovered the surprising fact that the path-integral quantization
of a scalar particle moving in a Riemannian manifold $\cM$ leads to
a quantum Hamiltonian operator $H$ that besides the expected kinetic
term, which is proportional to the Laplace Beltrami operator,
included a term of the form $\hbar^2 R/12m$, where $R$ is the Ricci
scalar curvature of the manifold \cite{DeWitt-1957}. See also
\cite{cheng-1972,Ringwood-1976}.

The fact that the curvature term is proportional to  $\hbar^2$ is a
clear indication that it is a by-product of the quantization of the
associated classical system. The latter is defined by a classical
Hamiltonian of the form
    \be
    H_c=\frac{1}{2m}g^{ij}p_ip_j,
    \label{H-c}
    \ee
where $g^{ij}$ are the coefficient of the inverse of  the metric
tensor $\bg=(g_{ij})$ in a local coordinate frame, and  Einstein's
summation convention is employed.

In the canonical quantization program, the coefficient  of the
curvature term is related to the choice of ordering of factors in
the quantum analog of (\ref{H-c}). Indeed different factor-ordering
prescriptions that yield a scalar Hamiltonian operator $H$
correspond to different choices for the coefficient of the curvature
term; in general,
    \be
    H=-\frac{\hbar^2}{2m}\Delta_g+\frac{\lambda\hbar^2}{m}\,R,
    \label{H=}
    \ee
where $\Delta_g$ is the Laplace-Beltrami operator for  the metric
tensor $\bg$, i.e., the operator acting on the scalar functions
$\psi:\cM\to\R$ according to
    \be
    (\Delta_g\psi)(x):=g^{-1/2}\partial_i \left[g^{ij}
    g^{1/2}\partial_j\psi(x)\right],
    \label{Delta}
    \ee
$g:=\det(\bg)$, and $\lambda$ is a real coefficient  whose choice
cannot be fixed using basic principles of quantum mechanics.

In the path-integral quantization scheme, the coefficient of the
curvature term turns out to depend on the choice of the path
integral measure \cite{Ali-1996}.  For example, in \cite{DeWitt
1992} DeWitt uses a different choice of the measure that corresponds to
$\lambda=1/8$ (rather than  $\lambda=1/12$ of \cite{DeWitt-1957}.)
See also \cite{Foerster-1994}. The choice $\lambda=1/8$ turns out to
be consistent with the result obtained by taking the bosonic part of
a supersymmetric quantum Hamiltonian  used in the path-integral
proofs of the Atiyah-Singer index theorem
\cite{Mostafazadeh-1994a,Mostafazadeh-1994b} where supersymmetry
removes the factor-ordering ambiguity. Other choices have also been
considered and promoted in the literature. For example References
\cite{Penrose-1965,Ryan-2004} show that the requirement of conformal
invariance of H corresponds to taking $\lambda=(n-1)/8(n-1)$, where
n is the dimension of $M$. For n=2, this gives $\lambda=1/8$.
References \cite{DeWitt-Morette-1980,Kleinert-1990} present
arguments supporting the choice $\lambda=0$. The review article
\cite{Marinov-1980} provides a summary of the related developments
up to the year 1980.

The author of Ref.~\cite{Ali-1996} points out that the ambiguity
related to the choice of $\lambda$ could only be settled using the
experimental data obtained for the particular system in question.
This point of view was adopted independently in
Ref.~\cite{Mostafazadeh-1996} where a first step in this direction
was taken by computing  the effect of the scalar curvature term
$\lambda\hbar^2 R/m$ on the scattering cross section of a particle
moving in a cylindrically symmetric asymptotically flat surface.
Here the basic idea is to determine the dependence of the scattering
data on the value of $\lambda$ and try to pave the way for fixing
this value by comparing the theoretical results with the outcome of
a suitable scattering experiment. This is done by writing the
Hamiltonian operator (\ref{H=}) as the sum of the Hamiltonian
operator:
    \be
    H_0:=-\frac{\hbar^2}{2m}\, \nabla^2,
    \label{H-zero}
    \ee
for a free particle moving in a plane and an effective scattering
potential, namely
    \be
    V:=H-H_0.
    \label{V=}
    \ee
The latter is then treated as an perturbation, and the machinery  of
the first-order Born approximation is used to compute the scattering
amplitude and the cross section for $V$.

The preliminary results reported in \cite{Mostafazadeh-1996}
indicate that the scattering effect due to the geometry of a
Gaussian bump is actually not unrealistically small. However, there
is a basic difficulty with the experimental realization of these
results. This is because any realistic experimental setup that aims
at probing such an scattering effect would involve particles whose
motion is constrained to take place in a surface by certain
constraining forces that act in the three-dimensional Euclidean
space $\E^3$. In other words, the configuration space of the
particle is an embedded surface. It is well-known that the classical
mechanical system only involves the metric of the surface that is
induced by its embedding in $\E^3$. This does not carry any
information about the details of how the surface is embedded in
$\E^3$, i.e., it is only sensitive to the intrinsic geometry of the
surface. It is remarkable that the same does not seem to hold for a
quantum particle; a quantum particle would know about the extrinsic
geometry of the surface as well.

The study of the quantum mechanics of a particle constrained to
move in a manifold embedded in a Euclidean space has a long history.
There are two different approaches for dealing with this problem,
namely, Dirac's formulation of constrained Hamiltonian systems
\cite{Dirac-1964} and the thin-layer quantization scheme developed in
\cite{Jensen-1971,daCosta-1981,Tolar-1988,Maraner-1995,Ogawa-1992,Froese-2001,Schuster-2003}.

The application of Dirac's formulation to a particle constrained  to
move on a surface $S$ embedded in $\E^3$ yields second class
constraints whose details depend on the choice of the equation used
to characterize $S$. As different equations can describe the same
embedded surface, Dirac's method turns out to be ambiguous
\cite{Golovnev-2009}. See also \cite{Xun-2014}.

The thin-layer method as outlined in \cite{daCosta-1981} involves
three  steps. First, one considers a particle that is free to move
in a thin layer parallel to the embedded surface. Second, one
carries out a careful decoupling of the motion along the tangential
and normal directions to the embedded surface. Third, one uses a
careful limiting process that essentially removes the information
about the motion along the normal direction and yields a
Schr\"odinger equation in the tangential coordinates and a
corresponding effective Hamiltonian. This amounts to assuming that
the particle is in the ground state of a barrier potential that
keeps it in the vicinity of the surface along the normal direction
\cite{Schuster-2003}. The Hamiltonian obtained by the thin layer
method has the form \cite{daCosta-1981,Tolar-1988,Schuster-2003}:
    \bea
    H
    &=&-\frac{\hbar^2}{2m}\Delta_g+\frac{\hbar^2}{2m}(K-M^2),
    \label{H-embed}
    \eea
where 
$M$ and $K=R/2$ are respectively the mean and Gaussian curvatures of
the  surface $S$, \cite{docarmo}. Unlike the Gaussian curvature,
which is uniquely determined by the metric tensor of $S$, the mean
curvature is sensitive to the way $S$ is embedded in $\E^3$, i.e.,
it is a measure of the extrinsic geometry of $S$. Similar results
have also been obtained within the context of Dirac's method for
particular choices of the constraint equation that specify $S$,
\cite{Ogawa-1991}. The authors of \cite{Kaplan-1997} show that the
choice of the constraining forces which in practice have a finite
strength can lead to the addition of a term proportional to
$\hbar^2$ to the geometric potential. Therefore, similarly to the
Dirac's method, the thin-layer quantization scheme that involves
realistic constraining forces also suffers from ambiguities in the
choice of the Hamiltonian operator.

To the best of our knowledge, the only experimental study of the
predictions  of the thin-layer quantization method is the one
reported in \cite{Onoe-2012}, where the authors consider the effect
of the geometric potential on the electronic properties of certain
liquids. The physical implications of the geometric potentials have
also been studied in
\cite{Atanasov-2007,Ortix-2011,Silva-2013,Vadakkumbatt-2014,Pahlavani-2015}.

The purpose of the present article is to use the approach of
\cite{Mostafazadeh-1996} to explore the phenomenon of geometric
scattering for an asymptotically Euclidean embedded surface $S$.
Specifically, we consider the geometric scattering of a scalar
particle of mass $m$ whose motion in $S$ is described by the
Hamiltonian operator:
    \be
    H=-\frac{\hbar^2}{2m}\Delta_g+\frac{\hbar^2}{m}(\lambda_1K+\lambda_2M^2),
    \label{H-gen}
    \ee
where $\lambda_1$ and $\lambda_2$ are arbitrary real coefficients.

\section{Geometric scattering amplitude}

We begin our analysis by recalling the Lippmann-Schwinger equation
for a  Hamiltonian of the form $H=H_0+V$,
    \begin{equation}
    |\psi^{(\pm)}\rangle=|\phi\rangle+\frac{1}{E- H_0\pm i\epsilon}\,
     V|\psi^{(\pm)}\rangle\;,
     \label{L-S}
     \end{equation}
where we use the notation of \cite{Sakurai}. In particular, $H_0$ is
the  free Hamiltonian, and $|\phi\kt$ and $E$ are respectively the
state vector and the energy of the incident particle that satisfy
$H_0|\phi\kt=E|\phi\kt$. The two-dimensional scattering problem for
the interaction potential $V$ consists of computing the scattering
amplitude $f(\vec k',\vec k)$ which is related to $|\psi^{(+)}\kt$
according to
    \begin{equation}
    \langle \vec x|\psi^{(+)}\rangle=\frac{1}{2\pi}\left[ e^{i\vec k\cdot\vec x}+
    \frac{e^{ikr}}{\sqrt{r}}f(\vec k',\vec k)\right].
    \end{equation}
Here $\vec x=(x^1:=x,x^2:=y)$ marks the cartesian coordinates in
$\R^2$,  $\vec k$ and $\vec k'$ are respectively the wavevector for
the incident and scattered wave functions, $k:=|\vec
k|=\sqrt{2mE}/\hbar$, $r:=|\vec x|$, and  $\vec k'= k \vec x/r$.

We can express the scattering amplitude in terms of the interaction
potential via
    \be
    f(\vec k',\vec k)= \frac{\sqrt{2 \pi}me^{-3i\pi/4}}{\sqrt{k  }\hbar^2}
    \int d^2\vec x'\,e^{-i\vec k'.\vec x'}\langle \vec x'|\hat V|\psi^{(+)}\rangle,
    \label{f=}
    \ee
and compute the differential cross section using:
    \begin{equation}\label{cs}
    \frac{d\sigma(\vec k',\vec k)}{d\Omega}=|f(\vec k',\vec k)|^2\;.
    \end{equation}
To perform the first Born approximation, we replace the
$|\psi^{(+)}\kt$  appearing on the right-hand side of (\ref{f=}) by
the state vector $|\vec k\kt$ for the incident particle. This gives
    \begin{equation}
    f(\vec k',\vec k)\approx f^{(1)}(\vec k',\vec k)=
    \frac{-i\sqrt{2\pi}me^{-i\pi/4}}{\sqrt{k}\hbar^2}
    \int d\vec x^{'2}e^{-i\vec k'\cdot\vec x'}\langle \vec x'|\hat V|\vec k\rangle.
    \label{q2.6}
    \end{equation}

For the geometric scattering problem determined by the Hamiltonian
operator  (\ref{H-gen}), the free Hamiltonian $H_0$ and the
interaction potential $V$ are respectively given by (\ref{H-zero})
and (\ref{V=}). These relations together with (\ref{H-gen}) imply
     \begin{align}
     \langle \vec x'|\hat V|\vec k\rangle & =
     \frac{\hbar^2}{4\pi m}
     \left[
     (g_0^{ij}-g^{ij})
     \partial'_i\partial'_j-
     \frac{\partial'_i(\sqrt{g}g^{ij})}{\sqrt{g}}\partial'_j
    + 2(\lambda_1 K + \lambda_2 M^2)
     \right] e^{i\vec k\cdot\vec x'},
     \label{xVk}
     \end{align}
where $g_0^{ij}$ stands for the components of the inverse of the
Euclidean  metric tensor (which coincides with the Kronecker delta
symbol $\delta_{ij}$ when $x^{\prime i}$ label Cartesian coordinates),
$\partial_i'$ means partial derivation with respect to the
$x^{\prime i}$, the quantities $g_0^{ij}$, $g^{ij}$, $g$, $K$, and $M$ are evaluated at $\vec
x'$, and we have employed $\langle \vec x'|\vec k\rangle=e^{i\vec
k\cdot\vec x'}/2\pi$.

\section{Geometric scattering for  a cylindrically symmetric surface}
\label{Sec3}

Suppose that the surface $S$ is the graph of a smooth function of
the radial  coordinate $r$ in the polar coordinate system in $\R^2$,
i.e., there is a smooth function $f:[0,\infty)\to\R$ such that
    \begin{equation}
    z=f(r)\; ,
    \label{cylindreqn}
    \end{equation}
where $(r,\theta,z)$ are cylindrical coordinates on $\mathbb{R}^3$.
This  equation determines a smooth embedded surface provided that it
has a vanishing derivative at $r=0$. That is $\dot f(0)= 0$, where
an overdot means derivation with respect to $r$.

We can identify $(r,\theta)$ with the polar coordinates and express
the  metric tensor induced from the Euclidean geometry of $\E^3$ on
$S$ in these coordinates as
    \begin{equation}
    [g_{ij}]=\left[
    \begin{array}{cc}
    1+ \dot{f}^2&0\\
    0&r^2
    \end{array}\right].
    \label{gcylindr}
    \end{equation}
Here the values $1$ and $2$ of the coordinate labels $i$ and $j$
correspond  to $r$ and $\theta$, respectively. In view of
(\ref{gcylindr}),  the Gaussian and mean curvatures of $S$ are
respectively given by
    \begin{align}
    &K=
    \frac{G \dG}{r},
    && M=
    \frac{1}{2}\left(\frac{G}{r}+ \dG \right),
    \label{KMcylindr}
    \end{align}
where
    \be
    G:=\frac{\df}{\sqrt{1+\df^2}}.
    \label{G=def}
    \ee
According to (\ref{KMcylindr}) and (\ref{G=def}), $K$ and $M$ are
regular  (nonsingular) functions of $r$ provided that  $\dot f(r)/r$
tend to a finite limit as $r\to 0$. In what follows we assume that
this condition holds.

In view of \eqref{xVk}, \eqref{gcylindr}, and \eqref{KMcylindr},
      \begin{align}
      \langle x'|\hat V|k\rangle
      &=\frac{\hbar^2}{4\pi m}\left\{
      -G^2\left(\frac{\vec k.\vec x'}{r^{'2}}\right)^2+
      i\left[\frac{1}{r'}G^2 + G \dG \right]\left(
      \frac{\vec k.\vec x'}{r'}\right)+ 2 \lambda_1 \frac{ G \dG}{r'} +
      \frac{\lambda_2}{2} \left( \frac{G^2}{r^2} + 2 \frac{G \dG}{r} + \dG ^2\right)
      \right\}e^{i\vec k.\vec x'}\;.
      \nonumber
      \end{align}
Substituting this relation in (\ref{q2.6}), we find
    \begin{eqnarray}
    \label{fkmaincyl}
    \nn f^{(1)}(\vec k',\vec k)&=&\frac{e^{-3\pi i/4}}{\sqrt{8\pi k}}
    \int d^2\vec x' e^{i(\vec k-\vec k').\vec x'} \left\{
    -G^2\left(\frac{\vec k.\vec x'}{r^{'}}\right)^2+
    i\left[\frac{1}{r'}G^2 + G \dG \right]\left(
    \frac{\vec k.\vec x'}{r'}\right) \right. \\
    &&\left. + 2 \lambda_1 \frac{ G \dG}{r'} + \frac{\lambda_2}{2}
    \left( \frac{G^2}{r^2}   + 2 \frac{G \dG}{r} + \dG ^2\right)
    \right\}e^{i\vec k.\vec x'}\;.
    \end{eqnarray}
In order to evaluate the integral in this equation, we work in a
Cartesian  coordinate system $(x',y')$ where  $\Delta\vec k:=\vec
k-\vec k'$ is along the $x'-$axis. Transforming to the corresponding
polar coordinates $(r',\theta')$ we can perform the integral over
$\theta'$. This gives
    \begin{eqnarray}
    f^{(1)}(\vec k',\vec k)&=&\sqrt{\frac{\pi}{2k}}e^{-3\pi i/4}
    \int_0^\infty dr' \left\{
    \left[-r'G^2k_x^2+ 2\lambda_1 G \dG +  \frac{\lambda_2}{2}
    \left( \frac{G^2}{r} +
    2 G \dG     + r \dG ^2\right)\right]J_0(r'|\Delta\vec k|) \right.\nonumber\\
    &&\left.+\left[-G^2 \frac{k_y^2-k_x^2}{|\Delta\vec k|}-k_x(G^2+
    r' G \dG )\right] J_1(r'|\Delta\vec k|)\right\}\;,
    \label{q91}
    \end{eqnarray}
where $J_0$ and $J_1$ are the Bessel functions of the first kind.
Denoting the  angle between  $\vec k$ and $\vec k'$ by $\Theta$, and
recalling that  $\vec k'= k \vec x/|\vec x|$, we have
    \begin{align*}
    &|\Delta\vec k|=2k\sin\left(\mbox{$\frac{\Theta}{2}$}\right)=2k_x,
    &&k_y^2-k_x^2=k^2\cos\Theta.
    \end{align*}
With the help of these relations, we can write (\ref{q91}) in the form
    \begin{eqnarray}
    f^{(1)}(\vec k',\vec k)&=&\sqrt{\frac{\pi}{2k}}e^{-3\pi i/4}
    \int_0^\infty dr \Big\{
    \Big[ -k^2 r \sin^2 (\mbox{$\frac{\Theta}{2}$})  G^2 + 2\lambda_1 G \dG +
    \frac{\lambda_2}{2} \Big( \frac{G^2}{r} + 2 G \dG + r \dG ^2\Big)\Big]
    J_0(2 k r \sin\mbox{$\frac{\Theta}{2}$})+\nonumber\\
    &&\Big[- \frac{k \,G^2}{2\sin\mbox{$\frac{\Theta}{2}$}}
    -k r \sin(\mbox{$\frac{\Theta}{2}$})
    G \dG \Big] J_1(2 k r \sin\mbox{$\frac{\Theta}{2}$})\Big\}.
    \label{main}
    \end{eqnarray}

Next, suppose that
    \beq
    \lim_{r\rightarrow \infty} r  J_1(2 k r \sin\mbox{$\frac{\Theta}{2}$})G(r)^2=0,
    \label{assump1}
    \eq
which roughly speaking means that as $r\rightarrow \infty$,
$|G(r)|$ tends to  $0$ faster than $r^{-1/4}$. In view of
(\ref{assump1}), the fact that
    \[\lim_{r\to 0} \df(r)=\lim_{r\to\infty}\df(r)=0,\]
and various properties of the Bessel functions, we have managed to
express  (\ref{main}) in the form
    \begin{eqnarray}
    f^{(1)}(\vec k',\vec k)&=&\sqrt{\frac{\pi}{2k}}e^{-3\pi i/4}
    \int_0^\infty dr \left[
    \frac{\lambda_2}{2} \left( \frac{G^2}{r} + r \dG ^2\right)
    J_0(2 k r \sin\mbox{$\frac{\Theta}{2}$}) \right.\nonumber\\
    &&\left.+  k \sin(\mbox{$\frac{\Theta}{2}$}) G^2
    \left(- \frac{1}{2\sin^2 \Theta/2} +
    2 \lambda_1 + \lambda_2\right)
    J_1(2 k r \sin\mbox{$\frac{\Theta}{2}$})\right].
    \label{eq-z01}
    \end{eqnarray}
For the forward scattering ($\Theta=0$), this equation reduces to
    \begin{eqnarray}
    f^{(1)}(\vec k,\vec k)&=&\sqrt{\frac{\pi}{2k}}e^{-3\pi i/4}
    \int_0^\infty dr \left[\frac{\lambda_2}{2} \left( \frac{G^2}{r} +
    r \dG ^2\right)
    -\frac{k^2 }{2} r G^2 \right].
    \end{eqnarray}
In particular, the Gaussian curvature of the surface does not affect
the  forward scattering amplitude. In contrast the scattering
amplitude for backscattering ($\Theta=\pi$) depends on both mean and
Gaussian curvatures of the surface;      \begin{eqnarray}
    f^{(1)}(\vec k'=-\vec k,\vec k)&=&\sqrt{\frac{\pi}{2k}}e^{-3\pi i/4}
     \int_0^\infty dr \left[\frac{\lambda_2}{2} \left( \frac{G^2}{r}
     + r \dG ^2\right)
    J_0(2 k r) 
    +  k   \left(2 \lambda_1 + \lambda_2- \frac{1}{2}\right)
    G^2 J_1(2 k r )\right].
    \end{eqnarray}

\begin{figure}[ht]
    \begin{center}
    \includegraphics[scale=.65]{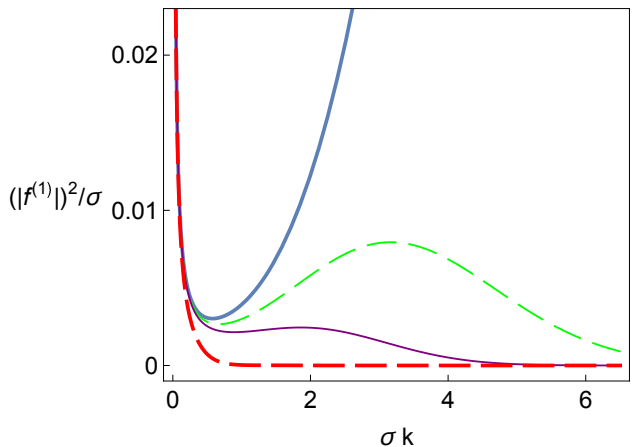}
    \caption{Plots of $|f^{(1)}|^2/\sigma$ as a function of $\sigma k$ for
    $\Theta=0$, i.e., forward scattering (thick solid blue curve),
    $\Theta=\pi/6$ (thin dashed green curve), $\Theta=\pi/4$ (thin solid
    purple curve), and $\Theta=\pi$, i.e., backward scattering (thick dashed
    red curve) for a Gaussian bump (\ref{gaussian}) with $\eta=0.1$. Here 
    we have taken $\lambda_1=-\lambda_2=\frac{1}{2}$ which follow from
    the thin-layer quantization scheme.}
    \label{fig1}
    \end{center}
    \end{figure}

As an example, consider a surface $S$ that has the shape of a
Gaussian  bump. Specifically, $S$ is given by (\ref{cylindreqn}) and
    \begin{equation}
    f(r)=\delta \, e^{- r^2/2\sigma^2},
    \label{gaussian}
    \end{equation}
where $\delta$ and $\sigma$ are real parameters. Let us introduce
the  dimensionless parameter:
    \[\eta:=\left(\frac{\delta}{\sigma}\right)^{\!2},\]
and compute the scattering amplitude (\ref{eq-z01}) as a power
series  in $\eta$. This gives
    \begin{align}
    \label{fkgaussian}
    f^{(1)}(\vec k',\vec k)&=\sqrt{\frac{\pi}{2k}}e^{-3\pi i/4}
    \left[\sigma^2k^2\left(\lambda_1\sin^2 \mbox{$\frac{\Theta}{2}$}
    -\frac{1}{4}\right)
    +\frac{\lambda_2}{4}\left(
    \sigma^4k^4 \sin^4 \mbox{$\frac{\Theta}{2}$}+2\right)\right]
    \exp\left(- \sigma^2k^2 \sin^2\mbox{$\frac{\Theta}{2}$} \right)\: \eta
    + \cO(\eta^2),
    \end{align}
where $\cO(\eta^\ell)$ stands for the terms of order $\ell$ and
higher in powers of $\eta$. For $|\eta|\ll 1$ we can safely ignore
$\cO(\eta^2)$, if $k$ is of the order of $\sigma^{-1}$ or smaller.
In particular,  for $\vec k'=\pm\vec k$,  we have
    \be
    f^{(1)}(\vec k',\vec k)=\frac{\eta}{4}\sqrt{\frac{\pi}{2k}}e^{-3\pi i/4}
    \times\left\{\begin{array}{ccc}
    2\lambda_2-(\sigma k)^2 & {\rm for} & \vec k'=\vec k,\\[12pt]
    e^{-(\sigma k)^2}\left[(4\lambda_1-1)(\sigma k)^2+\lambda_2[2+(\sigma k)^4]
    \right]
    &{\rm for} & \vec k'=-\vec k.\end{array}\right.
    \ee
This equation shows that we should be able to determine the
coefficients  $\lambda_1$ and $\lambda_2$ by examining the forward
and backward scattering data for incident particles with different
values of $k$. Figure~\ref{fig1} shows the plots of the differential
scattering cross section $|f^{(1)}(\vec k',\vec k)|^2$ for
$\Theta=0,\frac{\pi}{6},\frac{\pi}{4},\pi$ and the choice
$\lambda_1=-\lambda_2=\frac{1}{2}$ that is obtained in the
thin-layer quantization scheme \cite{daCosta-1981}.
According to this figure, there is a basic difference between the forward and non-forward scattering 
cross-sections. For $\Theta\neq 0$, the differential cross section has a peak that decreases in hight
and shifts to the left as we increase $\Theta$. 

\begin{figure}[ht]
    \begin{center}
    \includegraphics[scale=.65]{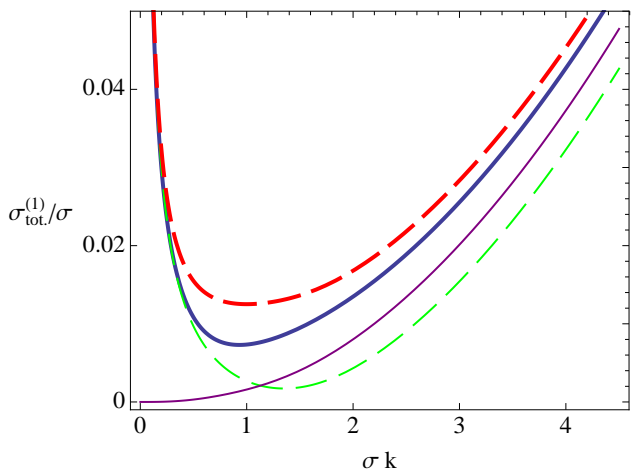}
    \caption{Plots of $\mbox{\large$\sigma^{(1)}_{\rm tot.}$}/\sigma$ as a
    function of $\sigma k$ for $\lambda_1=-\lambda_2=\frac{1}{2}$  (thick solid
    blue curve) which is obtained by the thin-layer quantization scheme
    \cite{daCosta-1981}, $\lambda_1=\lambda_2=\frac{1}{2}$ (thin dashed
    green curve), $\lambda_1-\frac{1}{2}=\lambda_2=0$ (thin solid purple curve),
    and $\lambda_2+\frac{1}{2}=\lambda_1=0$ (thick dashed red curve) for a
    Gaussian bump (\ref{gaussian}) with $\eta=0.1$}
    \label{fig2}
    \end{center}
    \end{figure}

We can also compute the total scattering cross section to leading
order in  $\eta$. The result is
    \be
    \mbox{\large$\sigma^{(1)}_{\rm tot.}$}=
    \int_0^{2\pi} |f^{(1)}(\vec k',\vec k)|^2d\theta=\frac{\pi^2}{256 k}
    e^{-\sigma^2k^2}\left[\fp_0(\sigma^2 k^2)I_0(\sigma^2k^2)+
    \fp_1(\sigma^2 k^2)I_1(\sigma^2k^2)\right]\eta^2+\cO(\eta^3),
    \label{sigma-tot=}
    \ee
where $I_n(\fz)$ stands for the modified Bessel function of the first kind, and
    \bea
    \fp_0(\fz)&:=&64\lambda_2^2+64\lambda_2(2\lambda_1-1) \fz+
    (16-64 \lambda_1 + 128\lambda_1^2 + 16\lambda_1\lambda_2 + 35 \lambda_2^2)\fz^2+\nn\\
    &&4 \lambda_2 (16 \lambda_1 + \lambda_2-4)\fz^3+8 \lambda_2^2 \fz^4,\nn\\[6pt]
    \fp_1(\fz)&:=&-2\left[(32\lambda_1^2+80\lambda_1\lambda_2+11\lambda_2^2)\fz+
    4(16\lambda_1^2+5\lambda_2^2+6\lambda_1\lambda_2-8\lambda_1-\lambda_2)\fz^2+
    \right.\nn\\
    &&\left.4\lambda_2(\lambda_2+8\lambda_1-2)\fz^3+4\lambda_2^2\fz^4\right].
    \nn
    \eea
Figure~\ref{fig2} shows the plots of the total scattering cross
section (\ref{sigma-tot=}) as a function of $\sigma k$ for different
choices of the parameters $\lambda_1$ and $\lambda_2$.

\section{Consequences of a small violation of cylindrical symmetry}

The results of the preceding section apply to surfaces with
cylindrical  symmetry. In this section we examine the effects of the
perturbations of the surface that violate this symmetry. We quantify
these by replacing (\ref{cylindreqn}) with
    \begin{equation}
    z=  f(r)+ \epsilon \sum_{n=1}^{\infty}  \left[a_n(r)\cos(n \theta)+
    b_n(r)\sin(n\theta)\right],
    \label{cylindreqnpert}
    \end{equation}
where $\epsilon$ is a real perturbation parameter,
$a_n,b_n:[0,\infty)\to{\mathbb{R}}$ are smooth functions that decay
asymptotically, i.e., $|a_n(r)|+|b_n(r)|\to 0$ as $r\to\infty$,
$\theta$ is the angular polar coordinate, and we demand that for all
$r\in[0,\infty)$,
    \[|\epsilon|\sum_{n=0}^\infty\Big[|a_n(r)|+|b_n(r)|\Big]\ll |f(r)|.\]
This allows us to ignore the quadratic and higher order terms in
powers of  $\epsilon$.

Equation~(\ref{cylindreqnpert}) defines an embedded surface that we
denote by  $\tilde S$. In cylindrical coordinates $(r,\theta) $, the
components of the metric $\tilde{g}$ of $\tilde S$ take the form:
    \[\tilde g_{ij} = g_{ij} + \epsilon\: {g_\epsilon}_{ij},\]
where $g_{ij}$ are given by \eqref{gcylindr}, and
    \begin{equation}
    [{g_\epsilon}_{ij}]:= \sum_{n=1}^{\infty} \left[
    \begin{array}{cc}
    2\dot{f} (\dot{a}_n  \cos n\theta +\dot{b}_n  \sin n\theta)
    & -n \dot{f}(a_n \sin n\theta- b_n \cos n\theta)   \\[6pt]
    -n \dot{f}(a_n \sin n\theta- b_n \cos n\theta)     & 0
    \end{array}
    \right]\;.
    \label{cylindrmetric}
    \end{equation}
Similarly, we write the corresponding Gaussian and mean curvatures
as
    \begin{align}
    & \tilde K= K + \epsilon\: K_\epsilon\;  && \tilde M= M +
    \epsilon\: M_\epsilon,
    \end{align}
where $K$ and $M$ are given by \eqref{KMcylindr}, and
    \begin{align}
    K_\epsilon :=&\sum_{n=1}^{\infty}  \cos (n \theta)
    \left( \frac{r \df \dda_n-n^2\ddf a_n  }{ r^2(1+    \df^2)^{2}}  +
    \frac{(1-3\df^2)\ddf\da_n }{r(1+\df^2)^{3}} \right) +
    \sum_{n=1}^{\infty}\sin (n  \theta)
    \left( \frac{r \df \ddb_n-n^2 \ddf b_n  }{ r^2(1+\df^2)^{2}}  +
    \frac{(1-3\df^2)\ddf \db_n }  {r(1+\df^2)^{3}}\right),
    \nn\\[6pt]
    M_\epsilon  := &\sum_{n=1}^{\infty}  \cos(n\theta)
    \left( -\frac{n^2 a_n }{2 r^2(1+\df^2)^{1/2}}
    -\frac{3\df \ddf \da_n }{2 (1+\df^2)^{5/2}} +
    \frac{  \da_n + r \dda_n}{2 r(1+\df^2)^{3/2}}\right)
    \nn\\
    &+ \sum_{n=1}^{\infty}  \sin (n\theta)
    \left(  -\frac{n^2 b_n }{2 r^2(1+\df^2)^{1/2}}
    -\frac{3\df \ddf \db_n }{2 (1+\df^2)^{5/2}} +\frac{  \db_n +
    r  \ddb_n}{2 r(1+\df^2)^{3/2}}\right)\; .
    \nn
    \end{align}
Recall that we require $r^{-1}\df(r)$ to tend to a finite value as
$r\to 0$, so that $K$ and $M$ do not have singularities. Similarly,
demanding $\tilde{K}$ and $\tilde{M}$ to be regular functions
restricts the choice of $f(r)$, $a_n(r)$, and $b_n(r)$.

We begin our analysis of the scattering of a scalar particle due to
nontrivial  geometry of $\tilde S$ by expressing the corresponding
scattering amplitude $\tilde{f}^{(1)}(\vec k',\vec k)$ in the form
    \be
    \tilde{f}^{(1)}(\vec k',\vec k)=f^{(1)}(\vec k',\vec k)
    +\epsilon f_\epsilon^{(1)}(\vec k',\vec k),
    \label{f=ff}
    \ee
where  $ f^{(1)}(\vec k',\vec k)$ is given by  \eqref{main}, and
$f_\epsilon^{(1)}(\vec k,\vec k)$ describes the effects of the
violation of cylindrical symmetry. To compute the latter, we employ
\eqref{q2.6} and the identities
    \bea
    \int_{0}^{2\pi} e^{i x \cos \theta} \cos (n \theta - \varphi) d\theta &=
    2i^n \pi J_n(x) \cos \varphi, \nn\\
    \int_{0}^{2\pi} e^{i x \cos \theta} \sin (n \theta - \varphi) d\theta &=
    -2i^n \pi J_n(x)\sin\varphi,\nn
    \eea
that hold for real variables $x$ and $\varphi$. The result of this calculation is
    \begin{align}
    f^{(1)}_\epsilon(\vec k',\vec k)&= \sqrt{\frac{\pi}{8k}}e^{-3\pi i/4}
    \sum_{\substack{n=-\infty \\ n\neq 0}}^{\infty}\int_{0}^{\infty} dr\;  i^n
    \Big\{ k^2
    \Big[\cos\Theta\:\cX[a_{|n|}(r)]+\frac{n}{|n|}\sin\Theta\:\cX[b_{|n|}(r)]\Big]
    J_{2+n}(2k r \sin (\Theta/2))\nn\\
    &+ k \Big[\sin(\Theta/2)\:\cY[a_{|n|}(r)]
     -\frac{n}{|n|}\cos (\Theta/2)\:\cY[b_{|n|}(r)]\Big]J_{1+n}(2k r \sin (\Theta/2))\nn\\
    &+ r \Big[2 \lambda_1 K_{|n|}^{(a)}
    + 4\lambda_2 M M_{|n|}^{(a)} - k^2\frac{\df \da_{|n|}}{ (1+\df)^2}\Big]
    J_n(2k r \sin (\Theta/2))\Big\},
    \label{fkkepsilon1}
    \end{align}
where $\cX$ and $\cY$ are differential operators that act on smooth
test  functions $\phi(r)$ according to
    \bea
    \cX[\phi(r)]&:=&\frac{n \df  \phi}{1+\df^2} -
    \frac{r \df \dot\phi}{(1+\df^2)^2},\nn\\
    \cY[\phi(r)]&:=&
    \frac{n(n+1) \df \phi }{r (1+\df^2)} +
    \frac{n \ddf \phi-2 \df \dot\phi -r \df \ddot\phi}{(1+\df^2)^2}
    -\frac{r \ddf (1-3 \df^2) \dot\phi}{(1+\df^2)^3},\nn
    \eea
and
    \begin{align*}
    K_{n}^{(a)} &  :=  \frac{r \df \dda_n-n^2\ddf a_n  }{ r^2(1+\df^2)^{2}}
    + \frac{(1-3\df^2)\ddf\da_n }{r(1+\df^2)^{3}} \; , \\
    M_{n}^{(a)} & :=   -\frac{n^2 a_n }{2 r^2(1+\df^2)^{1/2}}
    -\frac{3\df \ddf \da_n }{2 (1+\df^2)^{5/2}} +
    \frac{  \da_n + r \dda_n}{2 r(1+\df^2)^{3/2}}\;.
    \end{align*}

\begin{figure}[ht]
    \begin{center}
    \includegraphics[scale=.65]{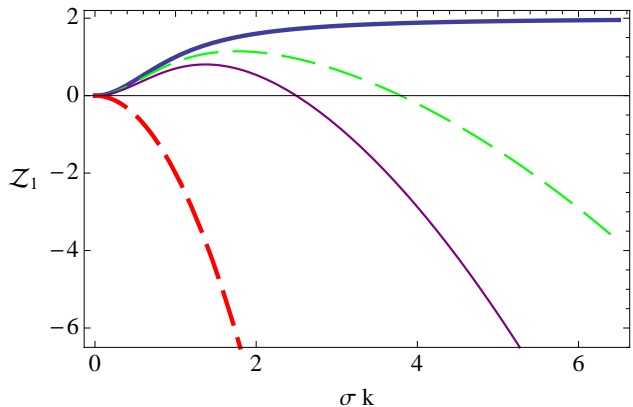}
    \caption{Plots of $\cZ_1$ as a function of $\sigma k$ for $\alpha_2=\sigma$ 
    and $\Theta=0$, i.e., forward scattering (thick solid blue curve), $\Theta=\pi/6$ 
    (thin dashed green curve), $\Theta=\pi/4$ (thin solid purple curve), and 
    $\Theta=\pi$, i.e., backward scattering (thick dashed red curve) for a perturbed 
    Gaussian bump (\ref{gaussian}) determined by (\ref{spec1}) and (\ref{spec2}) .}
    \label{fig3}
    \end{center}
    \end{figure}

For example, consider the perturbed Gaussian bump given by the
following choice for the functions $f$, $a_n$, and $b_n$:
    \begin{align}
    & f(r)=\delta \, e^{-r^2/2 \sigma^2},
    && a_1(r)=  \frac{r}{\alpha_1} f(r),
    &&a_2(r)=\frac{r^2}{\alpha_2^2} f(r),
    \label{spec1}\\
    & b_1(r)=  \frac{r}{\beta_1} f(r),
    && b_2(r)=\frac{r^2}{\beta_2^2} f(r),
    &&a_n(r) =b_n(r) =0,~\mbox{for $n\geq 2$},
    \label{spec2}
    \end{align}
where $\alpha_1$, $\alpha_2$, $\beta_1$ and $\beta_2$  are constant
parameters with the dimension of length. Then $f^{(1)}(\vec k',\vec
k)$ is given by (\ref{fkgaussian}), and we can evaluate the integral
on the right-hand side of  (\ref{fkkepsilon1}) to find:
    \begin{align}
    \label{fkgaussian1}
    \tilde{f}_\epsilon^{(1)}(\vec k',\vec k)=&\sqrt{\frac{\pi}{2k}}e^{-3\pi i/4}
    \Big\{
    - \frac{\sigma^4 k^2 \sin^2\mbox{$\frac{\Theta}{2}$}}{2\alpha_2^2}
    \left[\csc^2\mbox{$\frac{\Theta}{2}$}
    - \sigma^2 k^2  - 4 \lambda_1 \left(1-\sigma^2  k^2
    \sin^2\mbox{$\frac{\Theta}{2}$} \right)
    + \lambda_2 \sigma^4 k^4\sin^4 \mbox{$\frac{\Theta}{2}$}   \right] \\
    &+ \frac{i\sigma^2 k \sin(\mbox{$\frac{\Theta}{2}$})}{2\alpha_1 }
    \left[ - \sigma^2 k^2+
    4\lambda_1 \sigma^2 k^2\sin^2\mbox{$\frac{\Theta}{2}$}
    +\lambda_2\left(2+ \sigma^4 k^4
    \sin^4\mbox{$\frac{\Theta}{2}$}\right)\right] \Big\}
     \exp\left(-\sigma^2 k^2 \sin^2\mbox{$\frac{\Theta}{2}$}\right)\:
    \eta +  O(\eta^2).\nn
    \end{align}
Substituting (\ref{fkgaussian}) and  (\ref{fkgaussian1}) in
(\ref{f=ff}), we  can express the scattering amplitude for the
surface $\tilde S$ as
    \be
    \tilde{f}^{(1)}(\vec k',\vec k)={f}^{(1)}(\vec k',\vec k)\Big\{1+\epsilon\,
    \big[\cZ_1(\vec k',\vec k)+i\cZ_2(\vec k',\vec k)\big]\Big\},
    \label{tilde-f-1=}
    \ee
where
    \bea
    \cZ_1(\vec k',\vec k)&:=&\frac{2\sigma^2}{\alpha_2^2}
    \left[1-\sigma^2 k^2 \sin^2\mbox{$\frac{\Theta}{2}$}
    - \frac{ \lambda_2[(1-\sigma^2  k^2 \sin^2 \mbox{$\frac{\Theta}{2}$})^2+1]}{
    (4\lambda_1\sin^2\mbox{$\frac{\Theta}{2}$}-1)\sigma^2 k^2
    +\lambda_2\left(\sigma^4 k^4 \sin^4\mbox{$\frac{\Theta}{2}$}+2\right)}\right],
    \label{Z1=}\\[6pt]
    \cZ_2(\vec k',\vec k)&:=&\frac{2\sigma^2 k
    \sin\mbox{$\frac{\Theta}{2}$}}{\alpha_1 }.
    \label{Z2=}
    \eea
In particular, to the leading order in $\eta$ and $\epsilon$ the
differential  cross section for the surface $\tilde S$ has the form
    \be
    \frac{d\tilde\sigma^{(1)}(\vec k',\vec k)}{d\Omega}=
    |\tilde f^{(1)}(\vec k',\vec k)|^2=
    |f^{(1)}(\vec k',\vec k)|^2\left[1+2\epsilon\,\cZ_1(\vec k',\vec k)\right].
    \label{tilde-cross-sec=}
    \ee
According to this equation the effect of the violation of
cylindrical  symmetry that is given by (\ref{spec1}) and
(\ref{spec2}) is encoded in the value of $\cZ_1(\vec k',\vec k)$.
Figure~\ref{fig3} shows the graph of this quantity as a function of
$\sigma k$ for $\lambda_1=-\lambda_2=1/2$, which follow from
thin-layer quantization prescription, $\alpha_2=\sigma$, and
$\Theta=0,\pi/6,\pi/4$, and $\pi$. As seen from this figure we can
consistently apply (\ref{tilde-cross-sec=}) for values of $k$ that
are of the order of $\sigma^{-1}$ or smaller. In particular, for
$\sigma\epsilon/\alpha_1\ll 1$ and $\sigma\epsilon/\alpha_2\ll 1$,
we can ignore this kind of violations of cylindrical symmetry.
Notice also that for large values of $k$, the violation of cylindrical symmetry 
does not affect the forward scattering cross-section. This is not the case for 
non-forward scattering cross-section.

\section{Geometric scattering for a surface with local cylindrical symmetry}

Consider an embedded surface $\cS$ with local cylindrically
symmetric that is  given by
    \be
    z=\sum_{j=1}^N f_j(|\vec x-\vec \fc_j|),
    \label{local-cs}
    \ee
where $N$ is a positive integer, $f_j:[0,\infty)\to\R$ are smooth
functions such  that $\lim_{r\to 0}\dot f_j(r)/r$ exists, and
$\vec{\fc}_j=(a_j,b_j)$ are centers of local cylindrical symmetry.

Suppose that the functions $f_j$ decay sufficiently fast away from
$0$ so that for  each $j$ we can approximate $f_j(|\vec r-\vec
\fc_j|)$ by a function that vanishes outside a disc $\sD_j$ centered
at $\fc_j$ with $\sD_{j'}\cap\sD_j=\varnothing$ for $j'\neq j$. If
we use the first Born approximation to determine the geometric
scattering properties of such a surface, the scattering amplitude
for $\cS$ takes the form
    \be
    \ff^{(1)}(\vec k',\vec k)=\sum_{j=1}^N \ff_j^{(1)}(\vec k',\vec k),
    \label{sum-f}
    \ee
where $\ff_j^{(1)}(\vec k',\vec k)$ stands for the scattering
amplitude associated  with the surface $S_j$ given by
    \be
    z=f_j(|\vec r-\vec \fc_j|).
    \label{Sj}
    \ee
We can obtain $S_j$ from a surface $S_{0j}$ with cylindrical
symmetry about the  $z$-axis by a simple space translation. It is
not difficult to show that $\ff_j^{(1)}(\vec k',\vec k)$ is related
to the scattering amplitude  $f_{0j}^{(1)}(\vec k',\vec k)$ of
$S_{0j}$ according to
    \be
    \ff_j^{(1)}(\vec k',\vec k)=e^{i(\vec k-\vec k')\cdot\fc_j}
    f_{0j}^{(1)}(\vec k',\vec k).
    \label{trans}
    \ee
In view of (\ref{sum-f}) and (\ref{trans}), we can use the results
of  Sec.~\ref{Sec3} to compute the scattering amplitude of $\cS$.
This is particularly easy when $f_j$'s (and consequently
$f_{0j}^{(1)}(\vec k',\vec k)$'s) coincide. In this case,
    \be
    \ff^{(1)}(\vec k',\vec k)=C(\vec k',\vec k)f^{(1)}(\vec k',\vec k),
    \label{sum-f-id}
    \ee
where
    \be
    C(\vec k',\vec k):=\sum_{j=1}^N e^{i(\vec k-\vec k')\cdot\fc_j},
    \label{fF=}
    \ee
and $f^{(1)}(\vec k',\vec k)$ is the common value of
$f_{0j}^{(1)}(\vec k',\vec k)$.

If $\vec\fc_j$ form a lattice, $j$ stands for an index pair $(m,n)$ and
    \be
    \vec\fc_j=\vec\fc_{mn}=m\vec a+n\vec b,
    \label{cj=}
    \ee
where $\vec a$ and $\vec b$ are constant vectors. We can use this
relation to perform the sum in (\ref{fF=}). Supposing that $m$ and
$n$ respectively take  values in the intervals $[m_1,m_2]$ and
$[n_1,n_2]$, substituting (\ref{cj=}) in (\ref{fF=}), and using the
identity:
    \[\sum_{j=j_1}^{j_2}\fz^j=\frac{\fz^{j_2+1}-\fz^{j_1}}{\fz-1},\]
we obtain
    \be
    C(\vec k',\vec k)=\frac{(e^{i(m_2+1)k_a}-e^{im_1k_a})
    (e^{i(n_2+1)k_b}-e^{in_1k_a})}{(e^{ik_a}-1)(e^{ik_b}-1)},
    \label{C-lattice}
    \ee
where
    \begin{align}
    &k_a:=(\vec k-\vec k')\cdot\vec a,
    &&k_b:=(\vec k-\vec k')\cdot\vec b.
    \end{align}

A simple example is a finite lattice of Gaussian bumps, such as
those forming a  liquid-Helium Wigner lattice:
    \be
    z=\delta
    \sum_{m=m_1}^{m_2} \sum_{n=n_1}^{n_2} e^{-(\vec r-\vec\fc_{mn})^2/2\sigma^2},
    \label{gaussian-array}
    \ee
where $|\fc_{m'n'}-\fc_{mn}|\gg\sigma$ for $(m',n')\neq (m,n)$. For
this surface, the geometric scattering amplitude has the form
(\ref{sum-f-id}) with $C(\vec k',\vec k)$ and $f^{(1)}(\vec k',\vec
k)$ respectively given by (\ref{C-lattice}) and (\ref{fkgaussian}).
Notice however that in order for the quadratic and higher order
terms in $\eta$ on the right-hand side of (\ref{fkgaussian}) to be
negligible, we should have $(m_2-m_1)(n_2-n_1)\eta\ll 1$.

\begin{figure}[ht]
    \begin{center}
    \includegraphics[scale=.55]{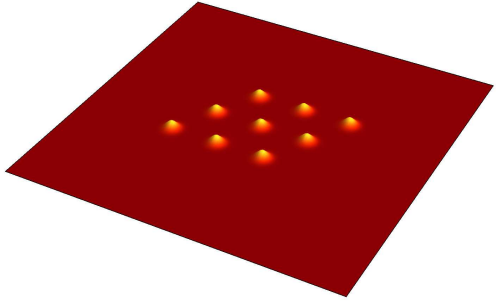}
    \caption{Schematic view of a surface involving the triangular
    lattice of Gaussian bumps given by (\ref{spec3}) and (\ref{spec4}).}
    \label{fig4}
    \end{center}
    \end{figure}

\begin{figure}[ht]
    \begin{center}
    \includegraphics[scale=.4]{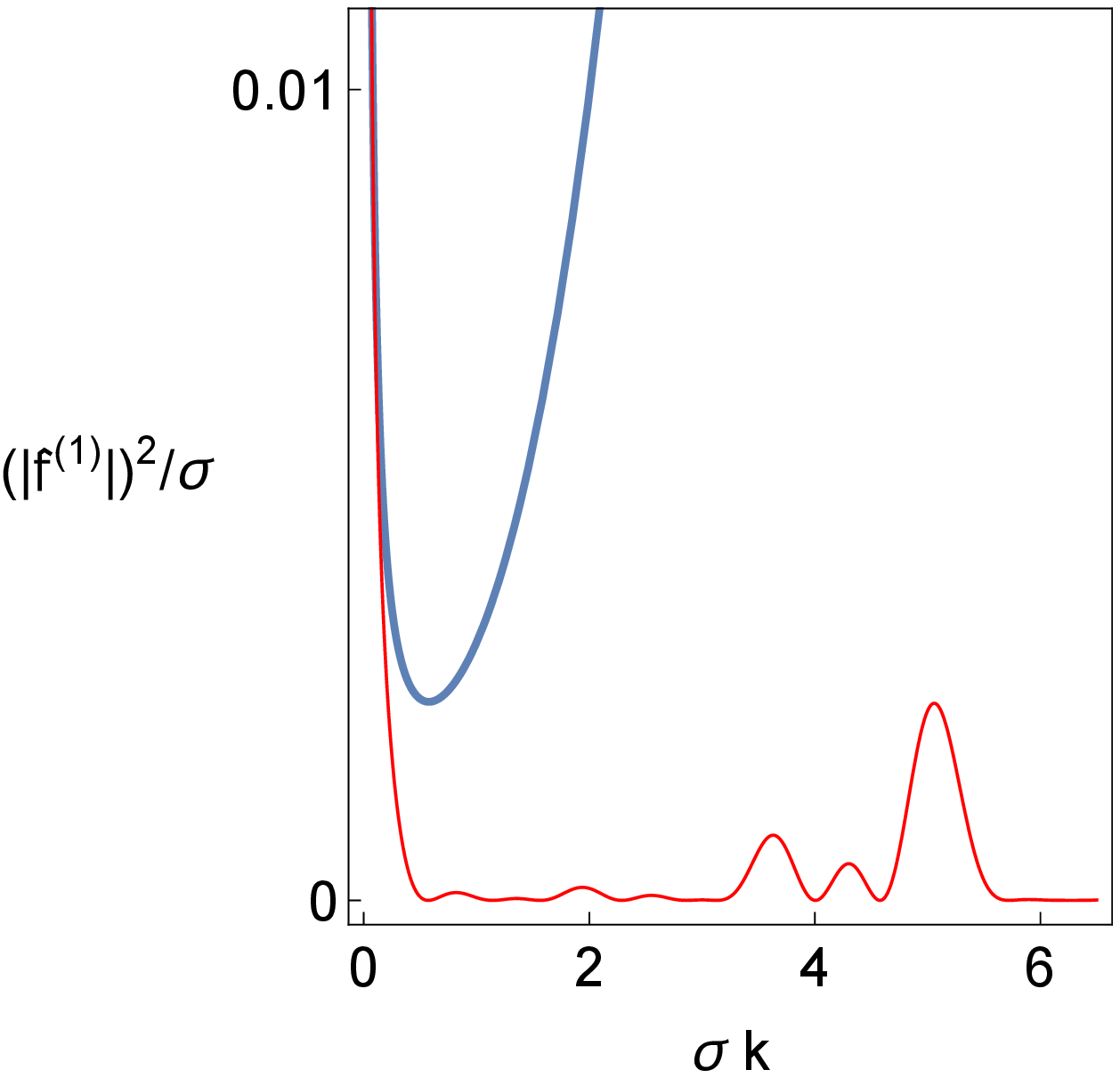}\hspace{1.5cm}
    \includegraphics[scale=.42]{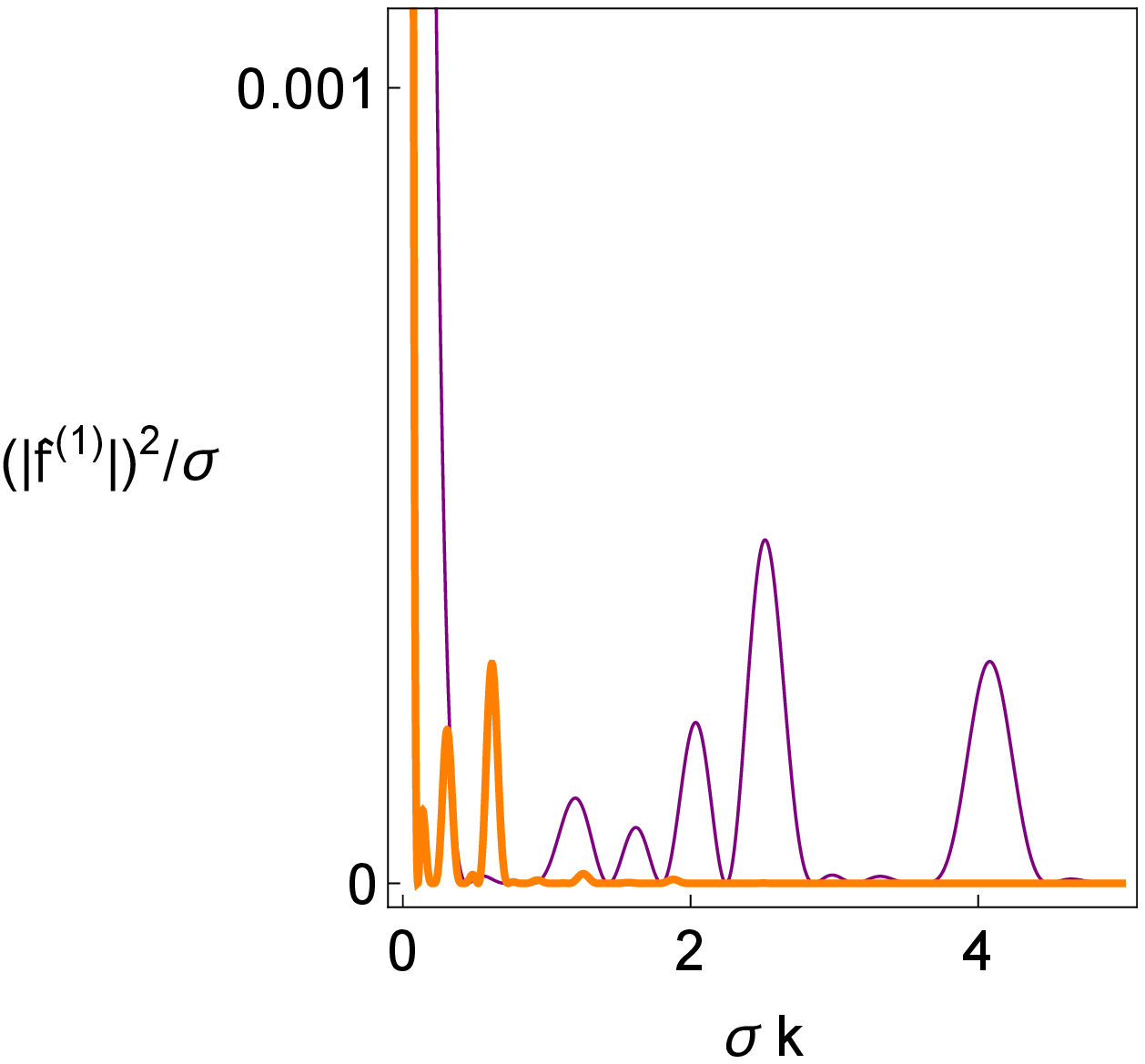}
    \caption{Plots of $|\ff^{(1)}(\vec k',\vec k)|^2/\sigma$ as a function of
    $\sigma k$ for the surface involving the triangular lattice of Gaussian
    bumps given by (\ref{cj=}), (\ref{gaussian-array}), (\ref{spec3}),
    (\ref{spec4}), $a=10\sigma$, $\eta=0.01$,
    $\lambda_1=-\lambda_2=1/2$, and $\Theta=0$ (left panel, thick blue curve),
    $\Theta=\pi/6$ (left panel, thin red curve), $\Theta=\pi/4$ (right panel,
    thin purple curve), and $\Theta=\pi$ (right panel, thick orange curve).}
    \label{fig5}
    \end{center}
    \end{figure}

To be specific, consider taking
    \begin{align}
    &\vec a=a(1,0), &&
    \vec b=a(\mbox{$\frac{1}{2},\frac{\sqrt{3}}{2}$}),
    \label{spec3}\\
    &m_1=n_1=-1, &&m_2=n_2=1,
    \label{spec4}
    \end{align}
where $a$ is the lattice constant. This corresponds to a triangular
lattice consisting of 9 Gaussian bumps, as depicted in
Fig.~\ref{fig4}.
In a coordinate system in which $\vec k$ is along the $x$-axis, we
have $\vec k'=k(\cos\Theta,\sin\Theta)$. This together with
(\ref{spec3}) imply
    \begin{align}
    & k_a=ak(1-\cos\Theta),
    &&k_b=\mbox{$\frac{ak}{2}$}(1-\cos\Theta-\sqrt 3\,\sin\Theta).
    \label{spec5}
    \end{align}
Substituting (\ref{spec4}) and (\ref{spec5}) in (\ref{C-lattice})
and using the  result together with  (\ref{fkgaussian}) and
(\ref{sum-f-id}) we can derive an analytic expression for the
geometric scattering amplitude of the surface defined by
(\ref{gaussian-array}). Figure~\ref{fig5} shows the graph of
differential cross section $|\ff^{(1)}(\vec k',\vec k)|^2$ as a
function of $\sigma k$ for $a=10\sigma$, $\eta=0.01$,
$\lambda_1=-\lambda_2=1/2$, and different values of $\Theta$.

\section{Summary and Concluding Remarks}

A classical free particle that is constrained to move on an embedded
surface feels the effect of the nontrivial geometry of the surface
through its contribution to the kinetic energy term in the
Hamiltonian. For a quantum particle there can be an additional
contribution to the Hamiltonian that arises in the form of a quantum
mechanical geometric potential involving both the Gaussian and mean
curvatures of the surface.  The strength of this curvature
interaction is determined by a pair of coupling constants whose
value cannot be determined from the first principles. For a specific
system these constants enter in the associated physical quantities. 

In this article we have considered a nonrelativistic spinless free
particle moving on an asymptotically flat embedded surface and
examined its scattering due to the nontrivial geometry of this
surface. In particular, we have used the first Born approximation to
calculate the geometric scattering amplitude for a surface with
global cylindrical symmetry and examined the effects of
perturbations of the surface that violate this symmetry. We have
also extended our analysis to surfaces with local cylindrical
symmetry. This allows for an analytic treatment of surfaces formed
by a finite lattice of well-separated bumps. Our results reveal the
possibility of determining the values of the unknown curvature
coefficients using the scattering data. 

For a cylindrically symmetric surfaces, only the mean curvature 
contributes to the forward scattering amplitude. This is not the case 
for the backward scattering amplitude that receive contributions 
from both the mean and Gaussian curvatures. In view of this 
observation, one can in principle determine the values of the unknown 
curvature constants only using the forward and backward scattering 
data. Therefore if it turns out that both $\lambda_1$ and $\lambda_2$ 
take nonzero values, then an experimental realization of our setup 
would provide means for independent measurements of the physical 
effects of intrinsic and extrinsic geometry of the surface.

Once the curvature coefficients are determined we can use our 
analytical results to make predictions on the behavior of the geometric 
scattering cross section and its dependence on the shape of the surface. 
For example, if the forward and backward scattering data for a Gaussian
bump confirm the choice given by the thin-layer quantization scheme, 
i.e., $\lambda_1=-\lambda_2=1/2$, we expect the differential 
cross-section for non-forward scattering to attain a single peak as
we vary the wavenumber of the incident wave. For a lattice of 
Gaussian bumps the differential cross-section develops several peaks.

\vspace{6pt} \noindent{\bf Acknowledgements.} This work has been
supported by the Scientific  and Technological Research Council of
Turkey (T\"UB{$\dot{\rm I}$}TAK) in the framework of the Project
No.~117F108 and by the Turkish Academy of Sciences (T\"UBA).

\end{document}